\documentclass[preprint,aps,eqsecnum]{revtex4}
\usepackage{graphicx}
\newcommand{\be}{\begin{eqnarray}}
\newcommand{\ee}{\end{eqnarray}}

\begin{document}
%\vspace*{1cm}
\title
{Parton Distributions in Impact Parameter Space}

\author{\bf H. Dahiya$^a$, A. Mukherjee$^b$, S. Ray$^b$}
%\email{dipankar@phys.ufl.edu}
\affiliation{$^a$ Department of Physics, Panjab University, 
Chandigarh 160014, India\\
$^b$ Department of Physics,
Indian Institute of Technology, Powai, Mumbai 400076,
India.}
%\date{\today\\[2cm]}
\date{\today}
\begin{abstract}
Fourier transform of the generalized parton distributions (GPDs) at zero
skewness with respect to the transverse momentum transfer gives the
distribution of partons in the impact parameter space. We investigate the
GPDs as well as the impact parameter dependent parton distributions
(ipdpdfs) by expressing them in terms of overlaps of light front wave
functions (LFWFs) and present a comparative study using three different 
model LFWFs. 
\end{abstract}
\maketitle

%%%%%%%%%%%%%%%%%%%%%%%%%%%%%%%%%%%%%%%%%%%%%%%%%%%%%%%%%%%%%%%%%%%%%
\section{Introduction}
%%%%%%%%%%%%%%%%%%%%%%%%%%%%%%%%%%%%%%%%%%%%%%%%%%%%%%%%%%%%%%%%%%%%%%
Deeply virtual Compton scattering (DVCS) $\gamma^*(q)+p(P) \rightarrow
\gamma(q')+p(P')$, where the virtuality of the initial photon $Q^2 =-q^2$ 
is much large compared to the squared momentum transfer $t=-(p-p')^2$, 
provides a valuable probe to the structure of the proton near the light 
cone. At leading twist, QCD factorization holds and DVCS amplitude can 
be expressed as a convolution in $x$ of the hard $\gamma^*q \rightarrow
\gamma q$ Compton amplitude with the generalized parton distributions
(GPDs) \cite{GPD}. Here $x$ is the light cone momentum fraction of the active 
quark. The skewness $\zeta = {Q^2\over 2 P \cdot q}$ measures the
longitudinal momentum transfer in the process.  

GPDs are richer in content about the hadron structure than ordinary 
parton distributions (pdfs). On one hand, $x$ moment of the GPDs 
give hadron form 
factors measurable in exclusive $e p \rightarrow e p $ scattering, on 
the other hand in the forward limit, i. e. for zero momentum transfer, 
they reduce to ordinary pdfs measurable in inclusive processes. Thus 
they provide a unified picture of the hadron.

GPDs are off-diagonal overlaps of light-front bilocal operators and unlike 
ordinary pdfs, they do not have an interpretation of probability densities. 
It has been shown in \cite{bur} that a Fourier transform 
(FT) of 
the GPDs with respect to the transverse momentum transfer $\Delta_\perp$ 
at zero skewness $\zeta$ gives the distribution of partons in the transverse 
position or impact parameter space. They are called impact parameter dependent 
parton distributions (ipdpdfs) $q(x,b_\perp)$.  The impact representation of pdfs
on the light front was first introduced by Soper \cite{soper} in the context
of the FT of the elastic form factor. $q(x,b_\perp)$ give 
simultaneous information on the
distribution of quarks as a function of $x$ and the transverse distance 
$b_\perp$ of the parton from the center of the proton in the transverse plane. 
Ipdpdfs obey certain positivity constraints and thus,
it is legitimate to physically interpret them as probability densities. 
In fact, 
this interpretation is not limited by relativistic effects in the 
infinite momentum 
frame. $q(x,b_\perp)$ are defined for a hadron state at sharp momentum 
$P^+$ localized in the transverse plane such that the transverse center of 
momentum is at $R_\perp=0$ (one can also work with a wave packet state 
localized in the transverse position space in order to avoid the state to 
be normalized to a delta function). When the target is transversely
polarized, the distribution of partons in the impact space is no longer
axially symmetric and the deformation is described by the FT of the GPD 
$E(x,0,t)$. This distortion has been shown to be connected with Sivers
effect \cite{metz,burk1}.  
A 3-D picture of the quarks and gluons in the proton has been proposed in
\cite{ji} in terms of reduced Wigner distributions which are related to the FT of the
GPDs in the rest frame of the proton. In \cite{skew} it has been shown that 
a FT in the skewness $\zeta$ provides a boost invariant longitudinal 
position space
picture of the proton and is analogous to optical diffraction obtained in
single slit experiment. Thus, GPDs provide a unique picture of
the hadron in transverse and longitudinal position space. 

GPDs can be expressed in the light-front gauge as overlaps of the
light-front wave functions (LFWFs) of
the target hadron. These are off-forward overlaps in general and one
requires not only a particle number conserving $n \to n $ overlap similar to
forward pdfs but also
an $n+1 \to n-1$ overlap, when the parton number is decreased by two. When
$\zeta$ is zero, the second contribution vanishes. GPDs and the ipdpdfs have
been investigated in several phenomenological models, for example in chiral
quark model for the pion \cite{model1}, in the constituent quark model
\cite{model2}, in terms of a
power law ansatz of the light cone wave function for the pion \cite{model3}, 
in the
context of investigating the color transparency phenomena \cite{model4}, 
in the
transverse lattice framework for the pion \cite{model5} and in the lattice 
framework \cite{lattice}.  

In this work we present a comparative study of the GPDs as well as the ipdpdfs using 
using several phenomenological models of hadron LFWFs. In \cite{orbit} 
DVCS amplitude at one loop in QED has been computed for a fermion. 
In effect, one
represents a spin ${1/2}$ system as a composite of a spin ${1/2}$ fermion
and a spin-$1$ vector boson, with arbitrary masses \cite{drell}. Similar models have been
used in \cite{dip,marc}. This one loop model is
self consistent since it has the correct correlation of different Fock
components of the state as given by the light-front eigenvalue equation.
In \cite{skew} a simulated model for the meson has been derived from the above LFWF,
by taking a derivative with respect to ({\it wrt}) the bound state mass square 
$M^2$, which appears
in the denominator of the wave function, thus improving the behaviour of it
near the end points at $x=0,1$. In this model, DVCS amplitude is purely
real. A similar power law behaviour has been used in \cite{power, model3} 
to construct the
GPDs for a meson. Differentiating once {\it wrt} $M^2$ generates a mesonlike
$k_\perp$ 
behaviour and differentiating {\it wrt} internal fermion mass $m^2$  and the
arbitrary gauge boson mass $\lambda^2$ simulates a protonlike behaviour of the
LFWFs. We present results in both these models. Models for LFWFs of
hadrons in $3+1$ dimensions displaying confinement at large distances and
conformal symmetry at short distances have been obtained using ADS/CFT
method. We also present ipdpdfs in this model.
%%%%%%%%%%%%%%%%%%%%%%%%%%%%%%%%%%%%%%%%%%%%%%%%%%%%%%%%%%%%%%%%%%%%%%%
\section{Generalized parton distributions}

The kinematics of DVCS process 
is given in detail in \cite{skew}. We work
in the frame of \cite{overlap}. We take skewness $\zeta$ to be zero. 
Momentum transfer is purely transverse; 
\be
t=(P-P')^2=-{\Delta_\perp}^2.
\ee
The generalized parton distributions
 $H$, $E$ are defined through matrix elements
of the bilinear vector currents on the light-cone:
\begin{eqnarray}
\lefteqn{
\int\frac{d y^-}{8\pi}\;e^{ix P^+y^-/2}\;
\langle P' | \bar\psi(0)\,\gamma^+\,\psi(y)\,|P\rangle
\Big|_{y^+=0, y_\perp=0}
} \hspace{2em}
\label{spd-def} \nonumber\\&=&
{1\over 2\bar P^+}\ {\bar U}(P') \left[ \,
H(x,\zeta,t)\ {\gamma^+}
 +
E(x,\zeta,t)\
{i\over 2M}\, {\sigma^{+\alpha}}(-\Delta_\alpha)
\right]  U(P)\ ,
\label{defhe}
\end{eqnarray}
here $\bar P={1\over 2}(P'+P)$ is the average momentum of the initial and
final hadron. 
 
The off-forward matrix elements can be expressed as overlaps of 
the light front wave functions \cite{overlap}. For non-zero skewness $\zeta$ 
there are diagonal parton number conserving contributions in 
the kinematical region $\zeta < x <1 $ and $\zeta-1 < x < 0$. 
There are 
off diagonal parton number changing contributions in the region $0<x<\zeta$.
In our case $\zeta=0$ and the only relevant kinematical region is
$0<x<1$.  
These correspond to the target helicity non-flip ($++$) and helicity flip
($+-$) contributions, respectively. If we consider a spin $1/2$ target state 
consisting of a spin $1$ particle and a spin $1/2$ particle, these
contributions can be expressed in terms of the $2$-particle LFWFs 
\cite{overlap},
\begin{eqnarray}
H_{(2\to 2)}(x,0,t)
&=&\int\frac{{\mathrm d}^2 {\vec k}_{\perp} }{16 \pi^3}
\Big[ \psi^{\uparrow *}_{+\frac{1}{2} +1}(x,{\vec k'}_{\perp})
\psi^{\uparrow}_{+\frac{1}{2} +1}(x,{\vec k}_{\perp})
+\psi^{\uparrow *}_{+\frac{1}{2} -1}(x,{\vec k'}_{\perp})
\psi^{\uparrow}_{+\frac{1}{2} -1}(x,{\vec k}_{\perp})
\nonumber\\
&&~~~~~~~~~~~~~~~~~~~~~~~~~~
+\psi^{\uparrow\ *}_{-\frac{1}{2} +1}(x,{\vec k'}_{\perp})
\psi^{\uparrow}_{-\frac{1}{2} +1}(x,{\vec k}_{\perp})
\Big] ,
\label{hnf}
\ee
\be
\lefteqn{
{(\Delta^1-{i} \Delta^2)\over 2M} E_{(2 \to 2)}(x,0,t)
}
\label{hf1} \nonumber\\
&=&
\int\frac{{\mathrm d}^2 {\vec k}_{\perp} }{16 \pi^3}
\Big[\psi^{\uparrow *}_{+\frac{1}{2} -1}(x,{\vec k'}_{\perp})
\psi^{\downarrow}_{+\frac{1}{2} -1}(x,{\vec k}_{\perp})
+\psi^{\uparrow *}_{-\frac{1}{2} +1}(x,{\vec k'}_{\perp})
\psi^{\downarrow}_{-\frac{1}{2} +1}(x,{\vec k}_{\perp})
\Big ] ,
\label{hf}
\end{eqnarray}
where
\begin{equation}
{\vec k'}_{\perp}={\vec k}^{~}_{\perp}-(1-x)\ {\vec{\Delta}}_{\perp}\ .
\label{xprime}
\end{equation}
Here $\psi^{\uparrow *}_{\lambda_1 \lambda_2}(x,{\vec k'}_{\perp})$ is the
lowest  (two-particle) Fock component of the hadron LFWF 
with helicity up and $\lambda_i$, $i=1,2$ are the
intrinsic helicities of the internal particles.  

The impact parameter dependent parton distributions are defined from the
GPDs by taking a FT in $\Delta_\perp$,
\be
q(x,b_\perp)= {1\over (2 \pi)^2} \int d^2 \Delta_\perp 
e^{-i b_\perp \cdot \Delta_\perp} H(x, t),\nonumber\\
e(x,b_\perp)= {1\over (2 \pi)^2} \int d^2 \Delta_\perp 
e^{-i b_\perp \cdot \Delta_\perp} E(x, t),
\label{ipd}
\ee
where $b_\perp$ is the impact parameter conjugate to $\Delta_\perp$. 
%%%%%%%%%%%%%%%%%%%%%%%%%%%%%%%%%%%%%%%%%%%%%%%%%%%%%%%%%%%%%%%%%%%%%%%%%%%%
\section{Simulated model calculations}
%%%%%%%%%%%%%%%%%%%%%%%%%%%%%%%%%%%%%%%%%%%%%%%%%%%%%%%%%%%%%%%%%%%%%%%%%%%%%
\begin{figure}[t]
\includegraphics[width=8cm,height=7cm,clip]{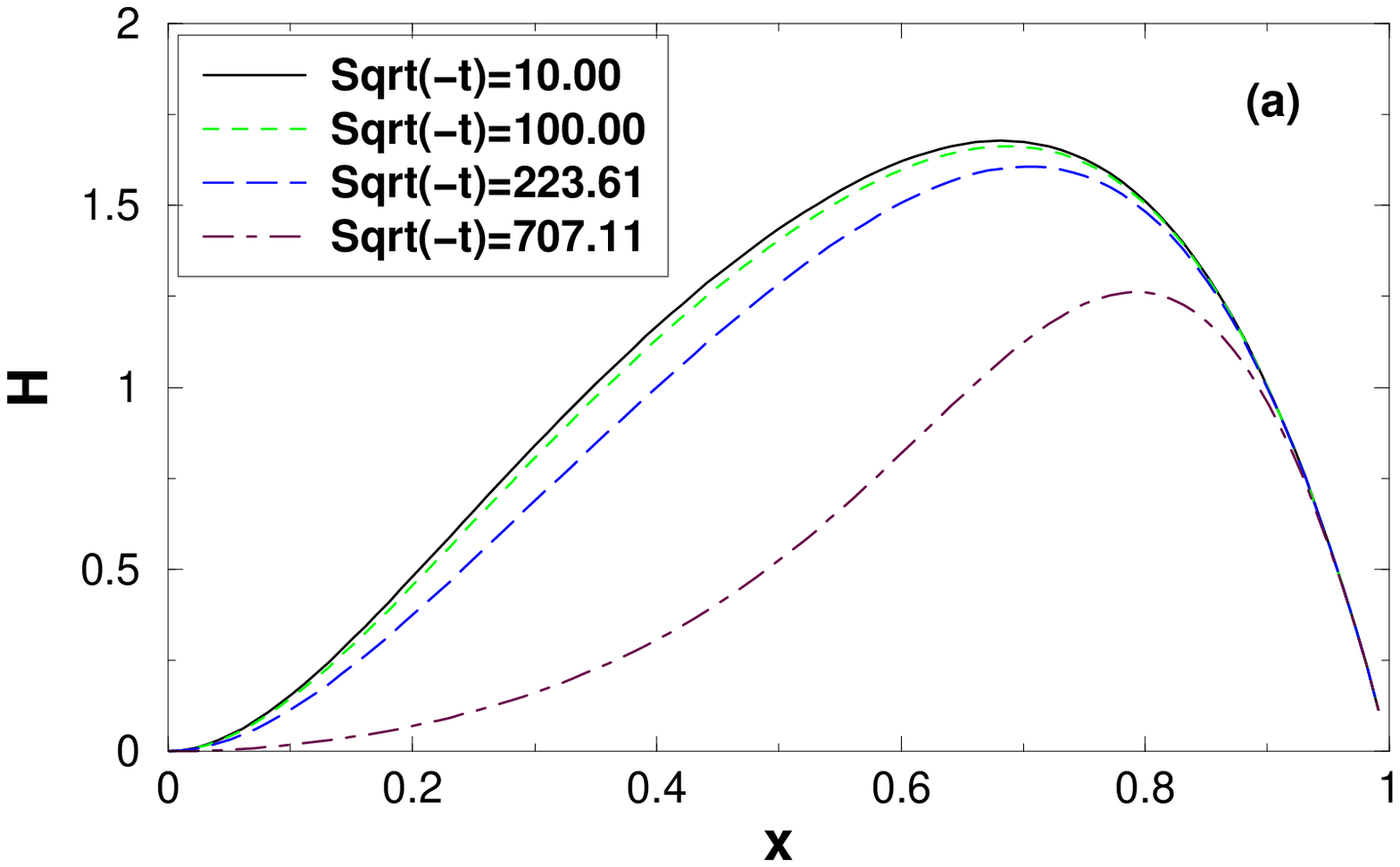}%
\hspace{0.2cm}%
\includegraphics[width=8cm,height=7cm,clip]{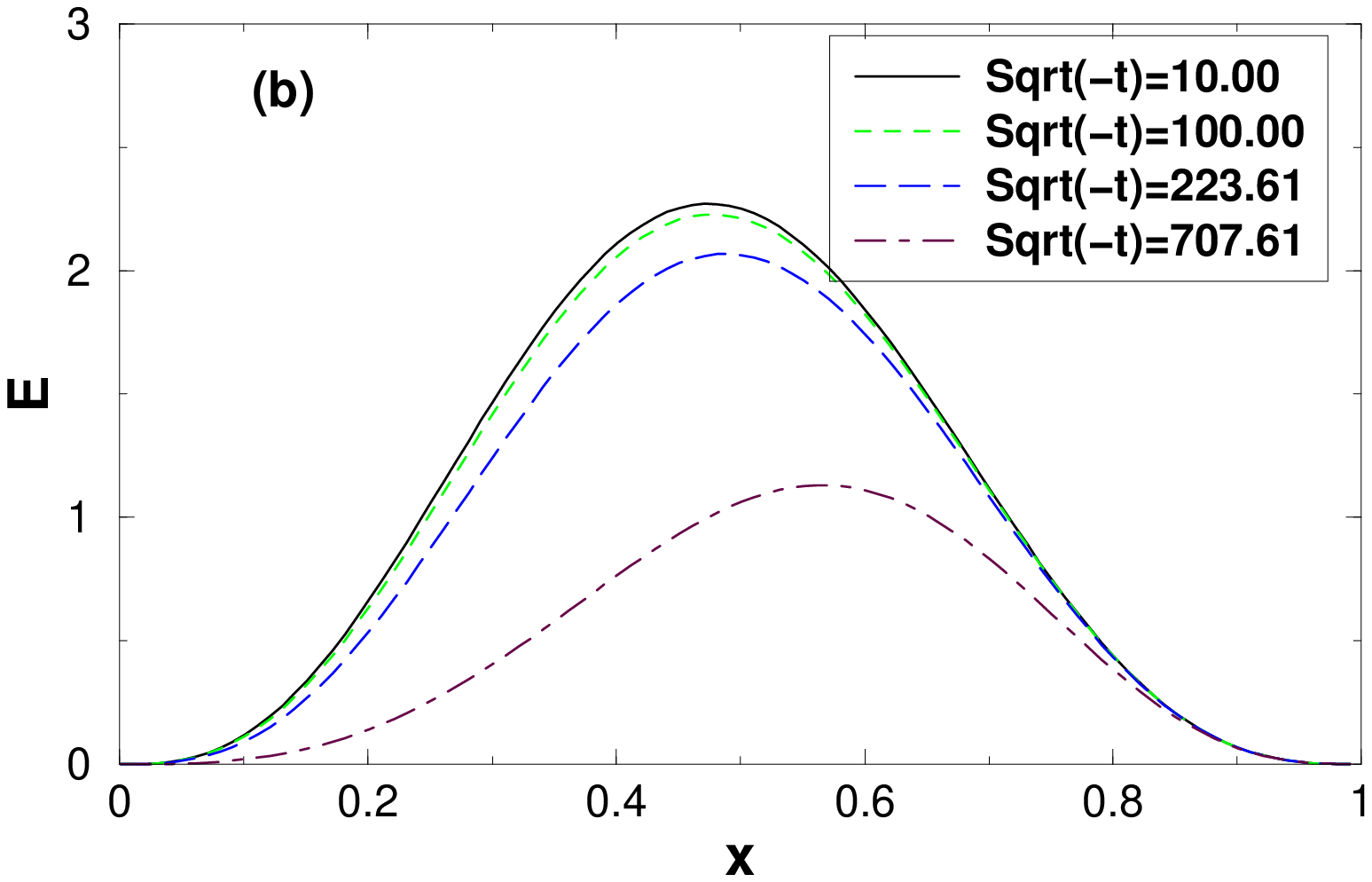}
\caption{\label{fig1} Generalized parton distributions in model $1$; (a)
$ H(x,0,t)$ and (b) $E(x,0,t)$ as a function of $x$ for fixed values of
$\sqrt{-t}$ in units of MeV.}
\end{figure}

In this section, we calculate the GPDs in simulated models of hadron LFWFs.
We start with the two-particle wave function for spin-up electron
\cite{orbit,drell,overlap}
\begin{equation}
\left
\{ \begin{array}{l}
\psi^{\uparrow}_{+\frac{1}{2}\, +1} (x,{\vec k}_{\perp})=-{\sqrt{2}}
\ \frac{-k^1+{i} k^2}{x(1-x)}\,
\varphi \ ,\\
\psi^{\uparrow}_{+\frac{1}{2}\, -1} (x,{\vec k}_{\perp})=-{\sqrt{2}}
\ \frac{k^1+{i} k^2}{1-x }\,
\varphi \ ,\\
\psi^{\uparrow}_{-\frac{1}{2}\, +1} (x,{\vec k}_{\perp})=-{\sqrt{2}}
\ (M-{m\over x})\,
\varphi \ ,\\
\psi^{\uparrow}_{-\frac{1}{2}\, -1} (x,{\vec k}_{\perp})=0\ ,
\end{array}
\right.
\label{vsn2}
\end{equation}
\begin{equation}
\varphi (x,{\vec k}_{\perp}) = \frac{e}{\sqrt{1-x}}\
\frac{1}{M^2-{{\vec k}_{\perp}^2+m^2 \over x}
-{{\vec k}_{\perp}^2+\lambda^2 \over 1-x}}\ .
\label{phi}
\end{equation}

Following the same references, we work in a generalized form of QED
by assigning a mass $M$ to the external electrons and a different
mass $m$ to the internal electron lines and a mass $\lambda$ to the
internal photon lines. The idea behind this is to model the
structure of a composite fermion state with mass $M$ by a fermion
and a vector `diquark' constituent with respective masses $m$ and
$\lambda$. Similarly, the wave function for an electron with negative 
helicity can also be obtained. In Eq. (\ref{phi}), the bound state
mass $M$ appears in the  energy denominator. As discussed in \cite{skew},
a differentiation of the QED LFWFs with respect to  $M^2$  improves the
convergence of the wave functions at the end points: $x=0,1,$ 
as well as improves the  $k^2_\perp$ behaviour,
thus simulating a bound state valence wavefunction. Differentiating once
with respect to $M^2$ will generate a meson-like behaviour of the LFWF.
However, this is not a model for a meson wavefunction
since the two constituents have spin half plus spin one.
If we differentiate once more we simulate the fall-off at short distances
which matches the fall-off wavefunction of a baryon, in the sense that the
form factor $F_1(Q^2)$ computed from the Drell-Yan-West formula will
fall-off like ${1\over Q^4}$. Here we have the analog of a two-parton
quark plus spin-one diquark model of a baryon, not three quarks.
Overlaps  of these wavefunctions in the same way as
have been done for the dressed electron wavefunctions will simulate the
corresponding GPDs. This is the approach we are following here.

\begin{figure}[t]
\includegraphics[width=8cm,height=7cm,clip]{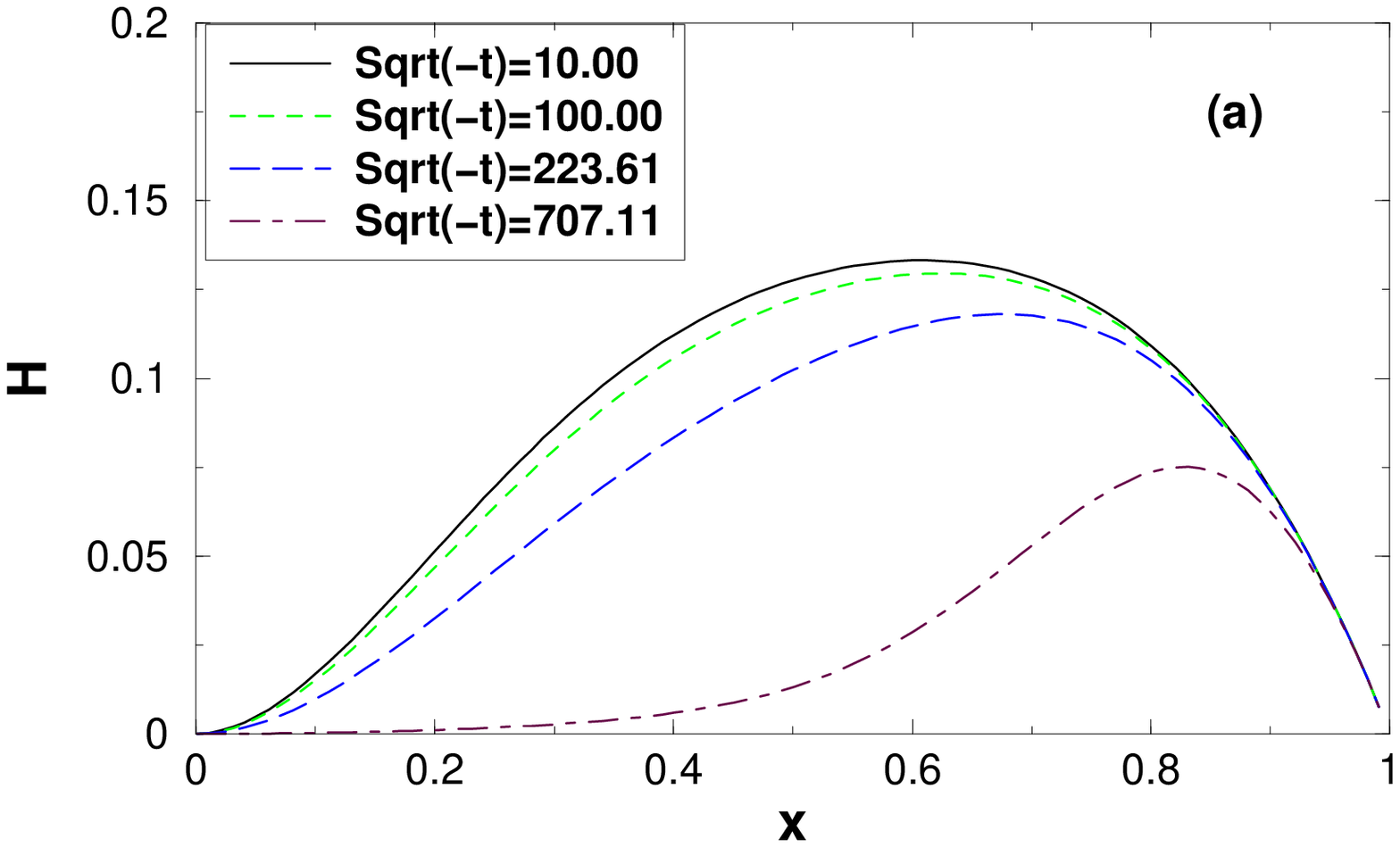}%
\hspace{0.2cm}%
\includegraphics[width=8cm,height=7cm,clip]{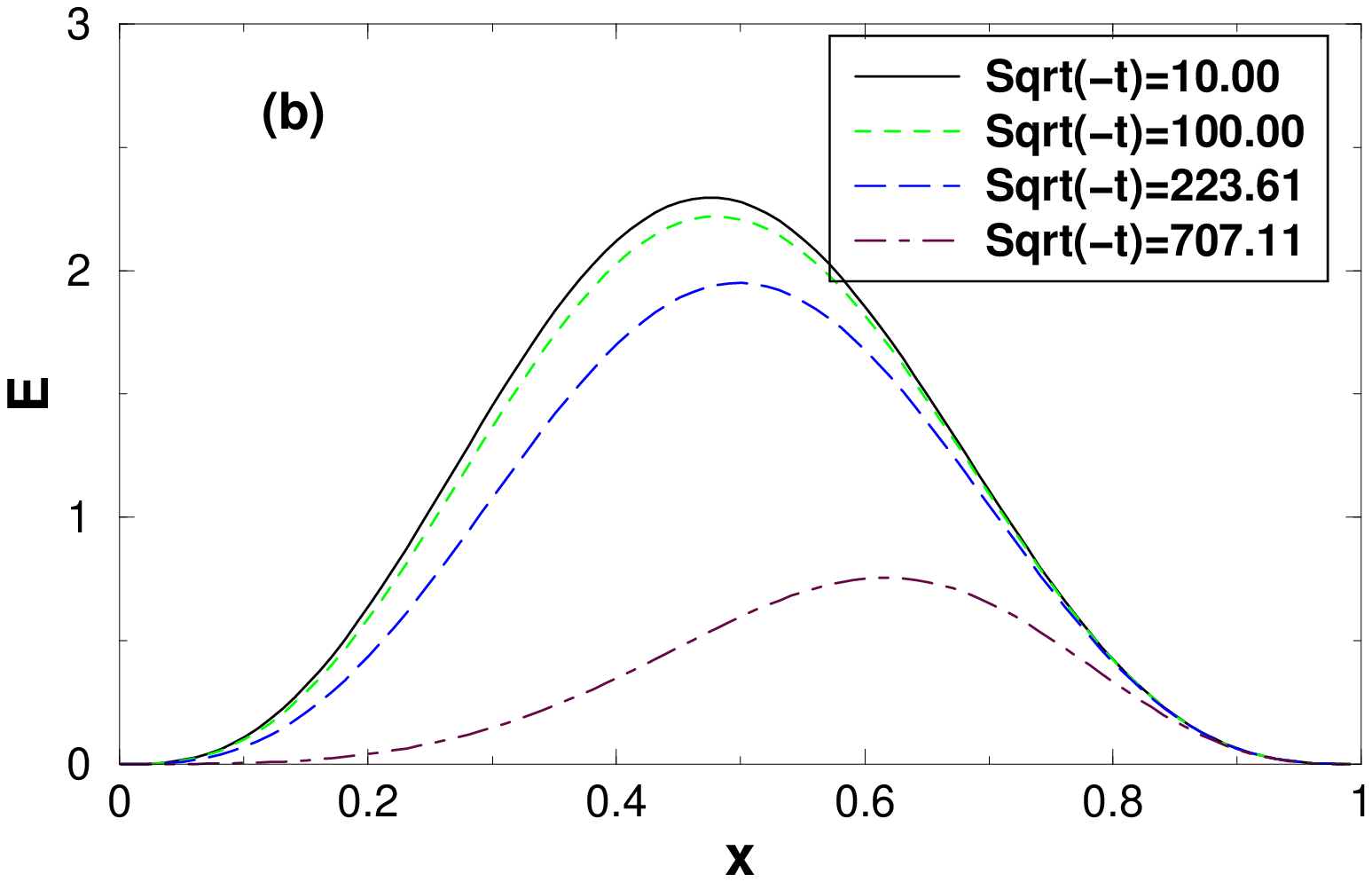}
\caption{\label{fig2} Generalized parton distributions in model $2$; (a)
$ H(x,0,t)$ and (b) $E(x,0,t)$ as a function of $x$ for fixed values of
${\sqrt{-t}}$ in units of MeV.}
\end{figure}

We differentiate $\varphi (x,{\vec k}_{\perp})$ with respect to the bound
state mass $M^2$ in order to simulate the LFWFs of a meson-like hadron. In
other words, we take
\begin{equation}
\varphi' (x,{\vec k}_{\perp}) = \mid {\partial \varphi (x,{\vec 
k}_{\perp})\over 
\partial M^2} \mid = N \frac{e}{\sqrt{1-x}}\
\frac{1}{\Big (M^2-{{\vec k}_{\perp}^2+m^2 \over x}
-{{\vec k}_{\perp}^2+\lambda^2 \over 1-x}\Big )^2}\ ;
\label{wfdenom}
\end{equation}
where $N$ is the normalization constant. We normalize the $2$- particle 
LFWF to $1$.

\begin{figure}[t]
\includegraphics[width=8cm,height=7cm,clip]{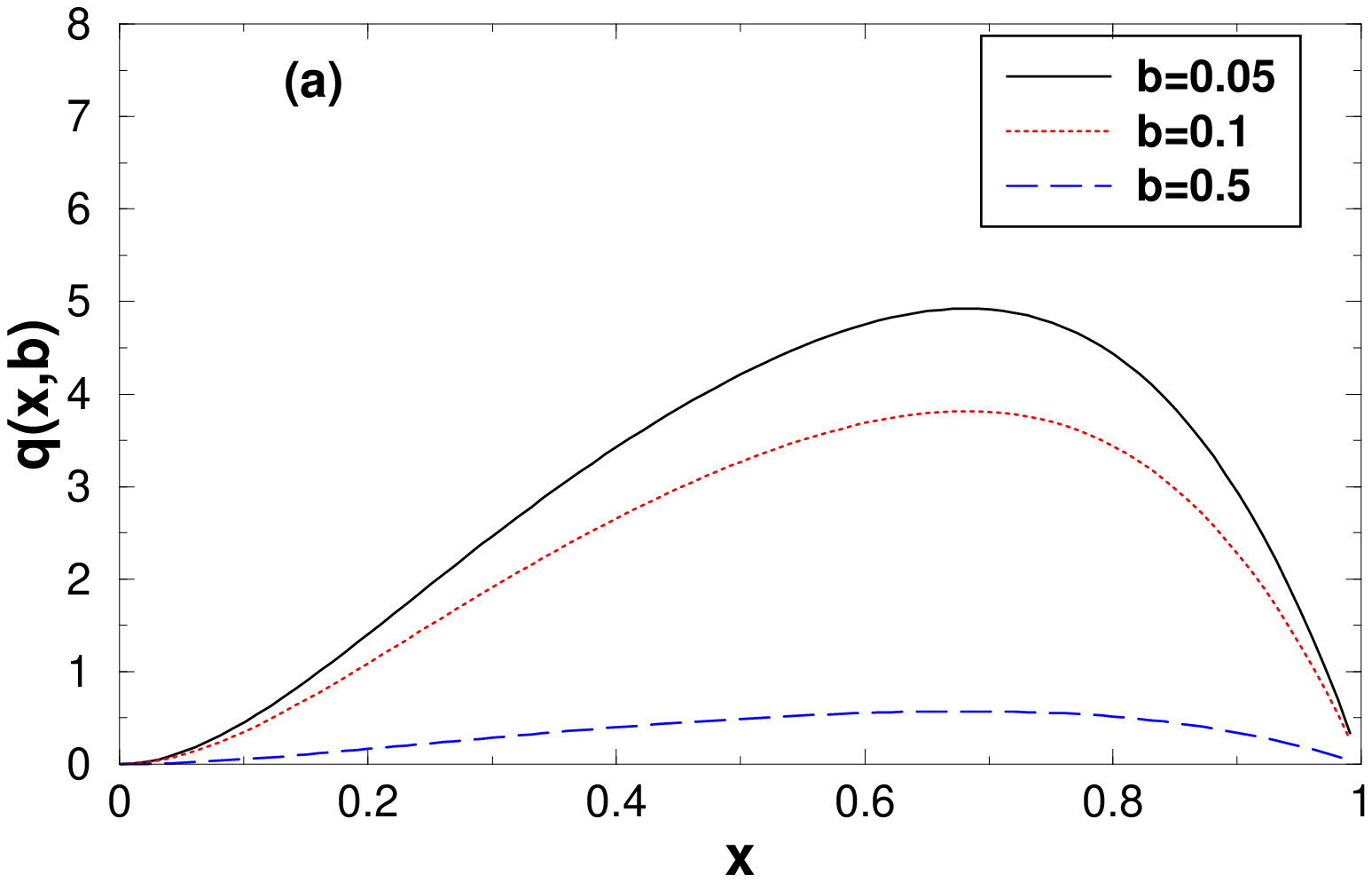}%
\hspace{0.2cm}%
\includegraphics[width=8cm,height=7cm,clip]{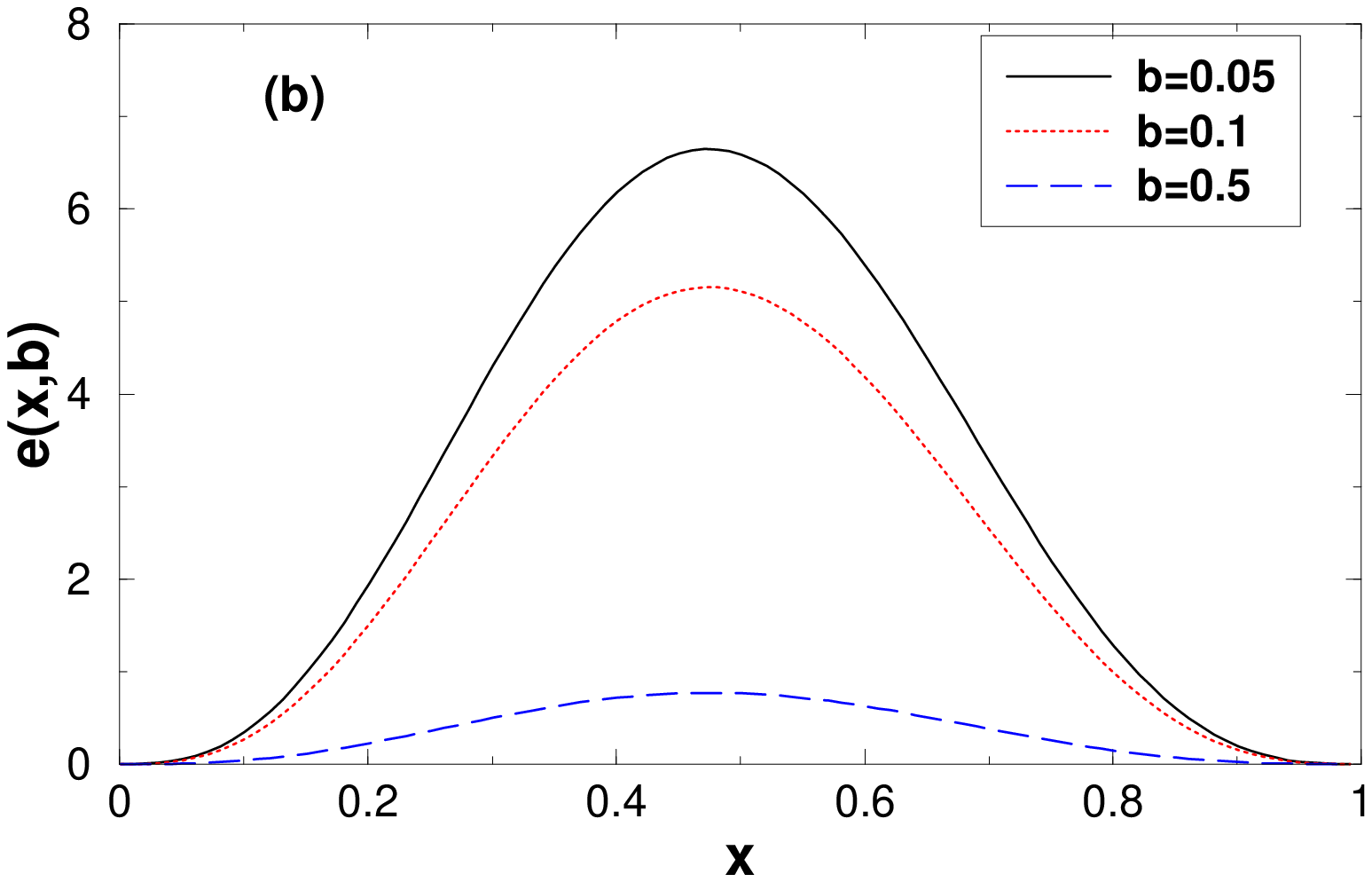}
\caption{\label{fig3} Impact parameter dependent parton distributions 
in model $1$; (a)
$ q(x,b_\perp)$ and (b) $e(x,b_\perp)$ as a function of $x$ for fixed 
values of $\mid b_\perp \mid $ in units of ${\mathrm{MeV}}^{-1}$}.
\end{figure}

The GPD in this model when the target helicity is not flipped (helicity non-flip) 
becomes (from Eq. (\ref{hnf})):
\be
H(x,t) &=& {e^2\over 16 \pi^3} N^2 x^2 (1-x) \Big [ \Big ( 
(1+x^2) I_1+(1+x^2) I_2 +\Big \{ (1+x^2) P \nonumber\\&&~~~~~~~~~~~+ 2
(1-x)^2 (M x-m)^2 I_3 \Big \} \Big ) \Big ],
\ee
where 
\be
I_1 &=& \int {d^2 k_\perp \over L_1^2 L_2} = 
\pi \int_0^1 d \beta {\beta \over D^2}\nonumber\\
I_2 &=& \int {d^2 k_\perp \over L_1 L_2^2}=\pi \int_0^1 d \beta {(1-\beta) 
\over D^2}\nonumber\\
I_3 &=& \int {d^2 k_\perp \over L_1^2 L_2^2}= 2 \pi \int_0^1 d \beta {\beta 
(1-\beta)\over D^3},
\ee
here $L_1=k_\perp^2+m^2 (1-x)+ \lambda^2 x -M^2 x (1-x)$, $L_2 = k_\perp^2 -
2 k_\perp \cdot \Delta_\perp (1-x) -B$ and 
\be
D =\beta m^2 (1-x) -\beta M^2 x (1-x) + \beta \lambda^2 x 
-(1-\beta) B-(1-\beta)^2 (1-x)^2 \Delta_\perp^2, 
\ee
where $B=M^2 x (1-x)-\Delta_\perp^2 (1-x)^2 -m^2 (1-x)- x \lambda^2$ and $P=
B+M^2 x (1-x) -m^2 (1-x) -\lambda^2 x $.

The helicity flip part as obtained from Eq. \ref{hf} is given as:
\be
E(x,t)= 2 M N^2 {e^2\over 8 \pi^3} x^3 (1-x)^3 ( M x -m )I_3.
\ee

Next, we simulate the LFWF for a proton-like hadron by differentiating eq.
(\ref{phi}) first {\it wrt} $m^2$ and then {\it wrt} $\lambda^2$. We get, 
\begin{equation}
\varphi ``(x,{\vec k}_{\perp}) = \mid { \partial \varphi 
(x,{\vec k}_{\perp})\over 
\partial m^2 \partial \lambda^2 } \mid = \frac{2 e}{x (1-x)^{3/2}}\
\frac{{\cal N}}{\Big (M^2-{{\vec k}_{\perp}^2+m^2 \over x}
-{{\vec k}_{\perp}^2+\lambda^2 \over 1-x}\Big )^3}\ .
\label{phi2}
\end{equation}
Here ${\cal N}$ is the normalization constant. As before, we normalize the
two particle LFWF to 1.

\begin{figure}[t]
\includegraphics[width=8cm,height=7cm,clip]{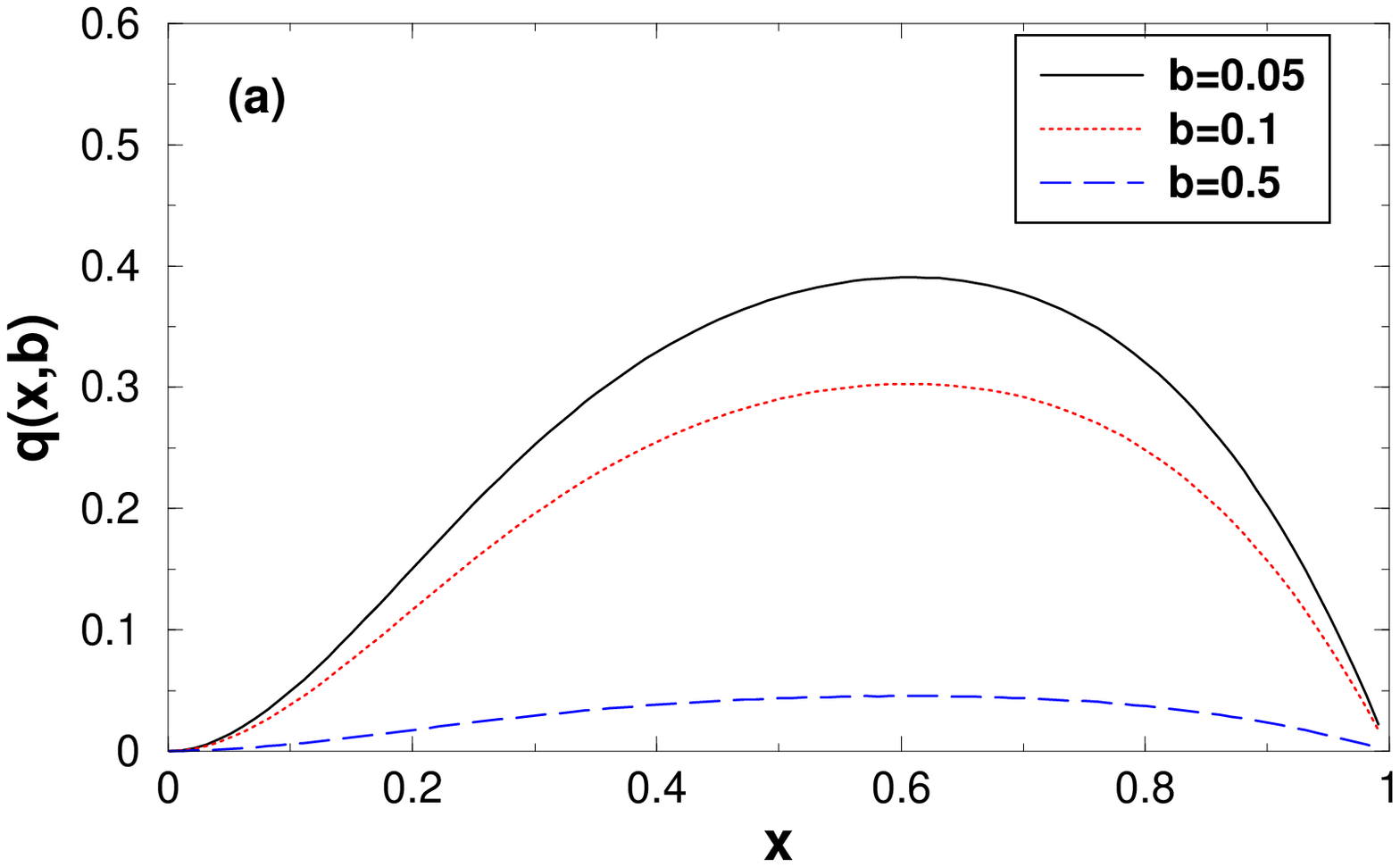}%
\hspace{0.2cm}%
\includegraphics[width=8cm,height=7cm,clip]{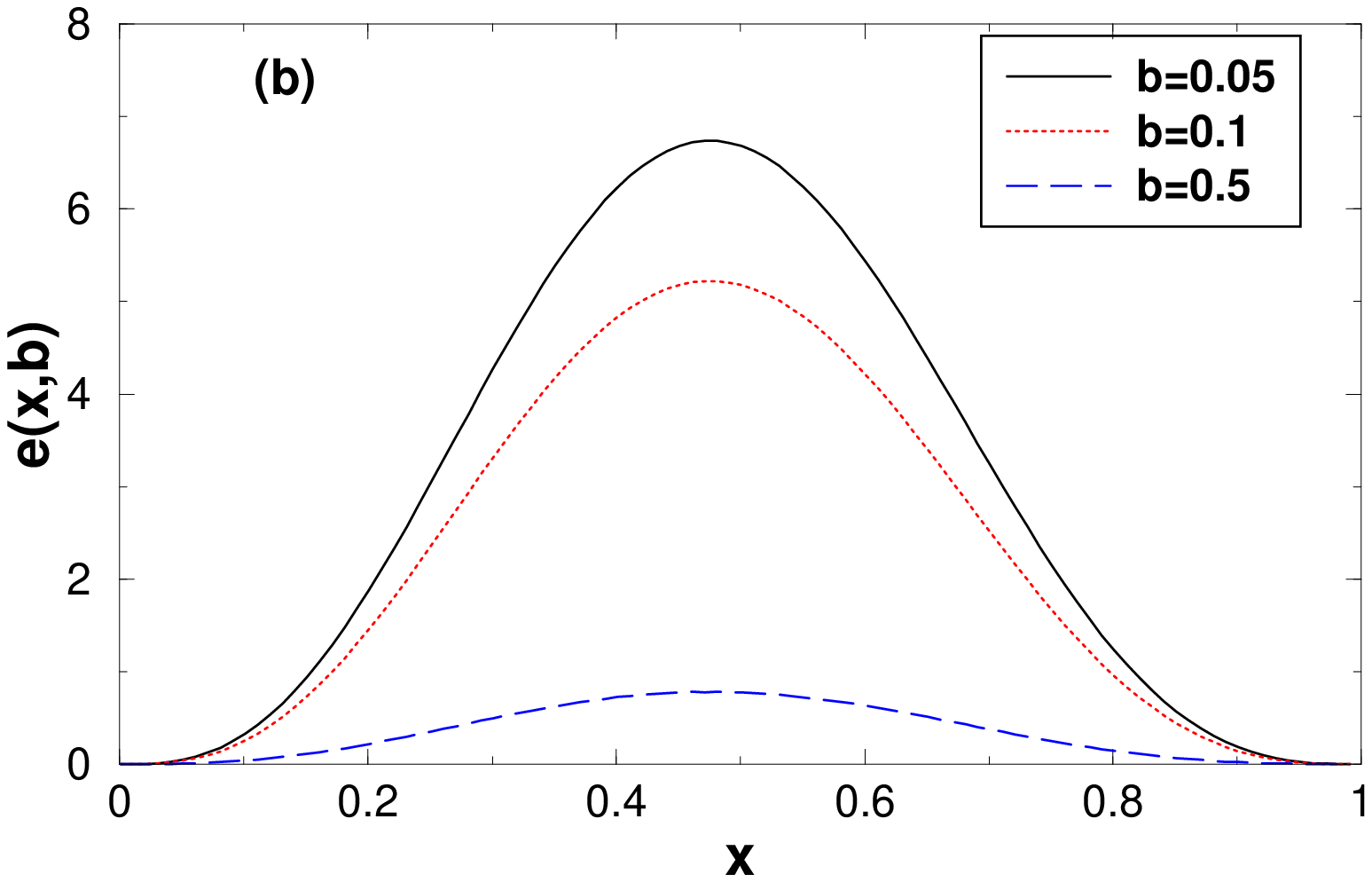}
\caption{\label{fig4} Impact parameter dependent parton distributions 
in model $2$; (a)
$ q(x,b_\perp)$ and (b) $e(x,b_\perp)$ as a function of $x$ for 
fixed values of $\mid b_\perp \mid $ in units of ${\mathrm{MeV}}^{-1}$.}
\end{figure}

The helicity non-flip GPD in this model becomes :
\be
H(x,t) &=& {e^2\over 4 \pi^3} {\cal N}^2 x^2 (1-x) \Big [ \Big ( 
(1+x^2) I'_1+(1+x^2) I'_2 
+\Big \{ (1+x^2) P \nonumber\\&&~~~~~~~~~~~~~~~~+ 2
(1-x)^2 (M x-m)^2 I'_3 \Big \} \Big ) \Big ],
\ee
where 
\be
I'_1 &=& \int {d^2k_\perp \over L_1^3 L_2^2}= 
3 \pi \int_0^1 d \beta {\beta^2 (1-\beta)\over D^4}\nonumber\\
I'_2 &=& \int {d^2k_\perp \over L_1^2 L_2^3}=
3 \pi \int_0^1 d \beta {(1-\beta)^2 \beta  \over D^4}\nonumber\\
I'_3 &=& \int {d^2k_\perp \over L_1^3 L_2^3}=
6 \pi \int_0^1 d \beta {\beta^2 (1-\beta)^2 \over D^5}.
\ee

The helicity flip part is given by,
\be
E(x,t)= 2 M {\cal N}^2 {e^2\over 2 \pi^3} x^3 (1-x)^3 ( M x -m )I'_3.
\ee

\begin{figure}[t]
\includegraphics[width=8cm,height=7cm,clip]{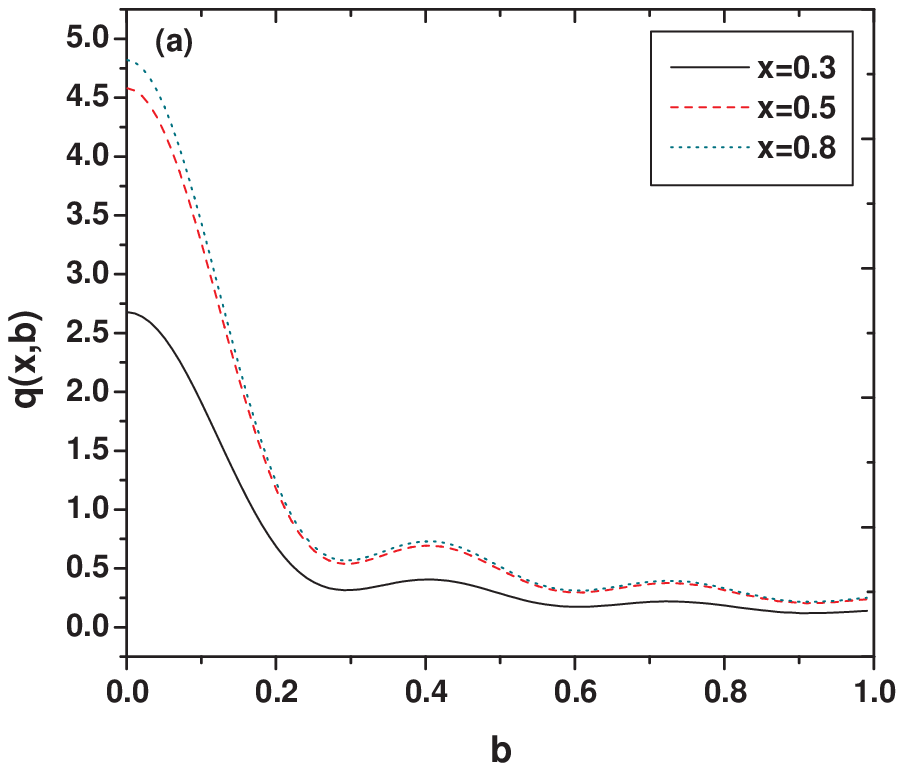}%
\hspace{0.2cm}%
\includegraphics[width=8cm,height=7cm,clip]{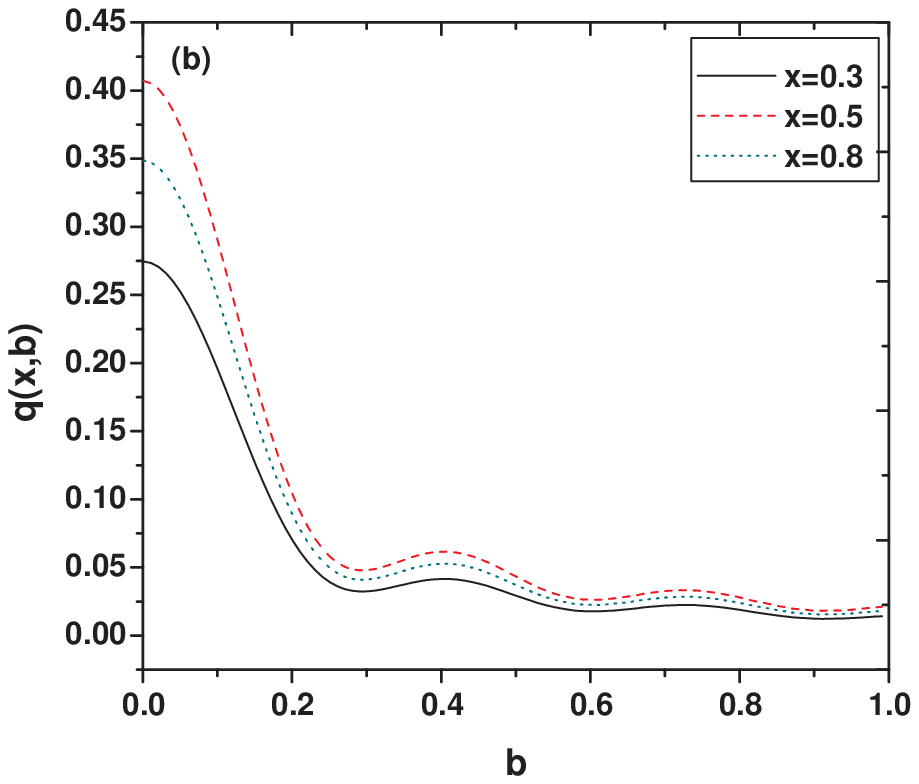}
\caption{\label{fig5} Impact parameter dependent parton distributions 
$q(x,b_\perp)$ in  (a) model 1 and (b) in model 2 as a function of 
$\mid b_\perp
\mid $ in units of ${\mathrm{MeV}}^{-1}$ for fixed values of
$x$.}
\end{figure}

Here we refer to the first model LFWF as model 1 and the second as model 2.
In both cases, we take $M=150$~ MeV, $m=\lambda=300$~ MeV \cite{skew}. 
In fig. 1 (a) we have plotted the helicity non flip GPD $H(x, 0, t)$ 
as a function of $x$ for fixed values of $t$. An interesting aspect of these
models is that in both of them one can generate a helicity flip
contribution $E(x,0,t)$, although the $k_\perp$ behaviour of model 1 is
similar to a meson. This is because, we have simulated this $k_\perp$
behaviour by using a two-particle composite state, where one of the
constituents has spin ${1/2}$ and the other has spin $1$. We have plotted 
the helicity flip GPD $E(x,0,t)$ in
model 1 in fig. 1 (b). In figs 2 (a) and (b) respectively, we have shown
$H(x,0,t)$ and $E(x,0,t)$ for model 2. From the figs, one can see 
that the qualitative behaviour 
of the GPDs in both models are similar. The GPD $H(x,0,t)$
increases with $x$, reaches a maximum and then falls to zero at $x \to 1$
independent of the momentum transfer $t$. As $x$ is the momentum
fraction of the active quark and at $x=1$ the active quark carries all the
momentum, contributions from the other partons are expected to be zero in this
limit and $H(x,0,t)$ is expected to become $t$ independent. The peak of $H$ 
occurs at higher values of $x$, and it shifts towards even 
higher values of $x$ as $\mid t \mid $ increases, which means that the 
active quark is more likely to have a larger momentum fraction. In fact,
here we are considering only the leading Fock space component of the target
hadron state. Higher Fock space components are more likely to contribute in
the small $x$ region. However,  
the magnitude of the peak decreases for high $\mid t \mid $, which is 
expected as the form factor decreases with $\mid t \mid $  for high $t$. In
model $2$, the peak shifts to larger $x$ faster with an increase of $\mid t
\mid $.      
The GPD $E(x,0,t)$ is associated with the amplitude when the nucleon helicity
flips and quark helicity does not flip. This is known to be related to the
quark orbital angular momentum. The functional  dependence of $E(x,0,t)$ is
different from $H(x,0,t)$ in both models. The peak is at $x=0.5$ for smaller
values of $\mid t \mid$ but shifts to larger values of $x$ as $\mid t \mid$
increases. It is zero at both $x=0,1$, same as $H(x,0,t)$.    
Like $H(x,0,t)$, it also becomes independent of $\mid t
\mid $ as $x \to 1$. $E(x,0,t)$ decreases as $\mid t
\mid$ increases, basically as the first moment of $E(x,0,t)$ gives the form
factor $F_2(t)$, which  decreases for large $\mid t\mid $. 

\begin{figure}[t]
\includegraphics[width=8cm,height=7cm,clip]{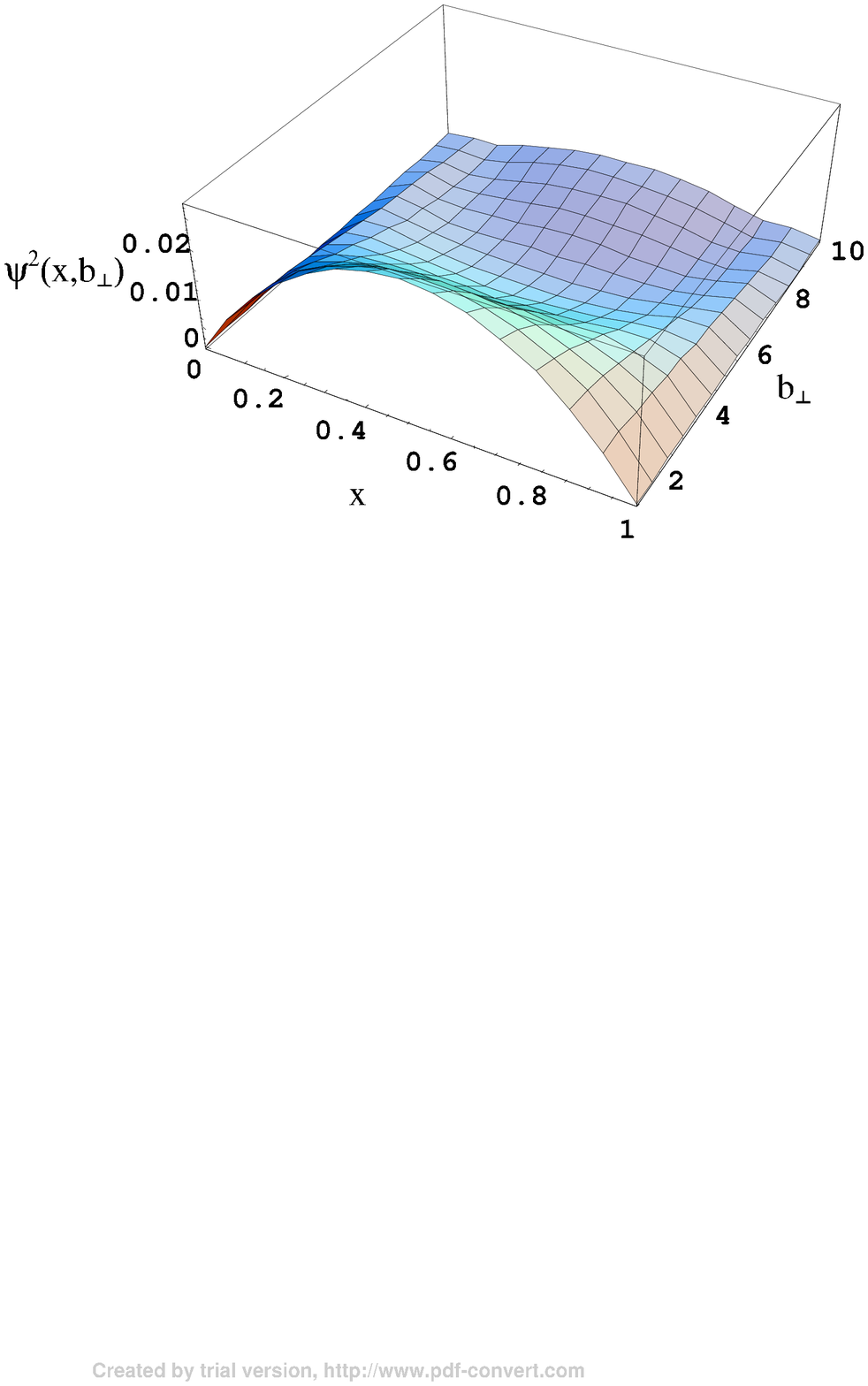}%
\hspace{0.2cm}%
\includegraphics[width=8cm,height=7cm,clip]{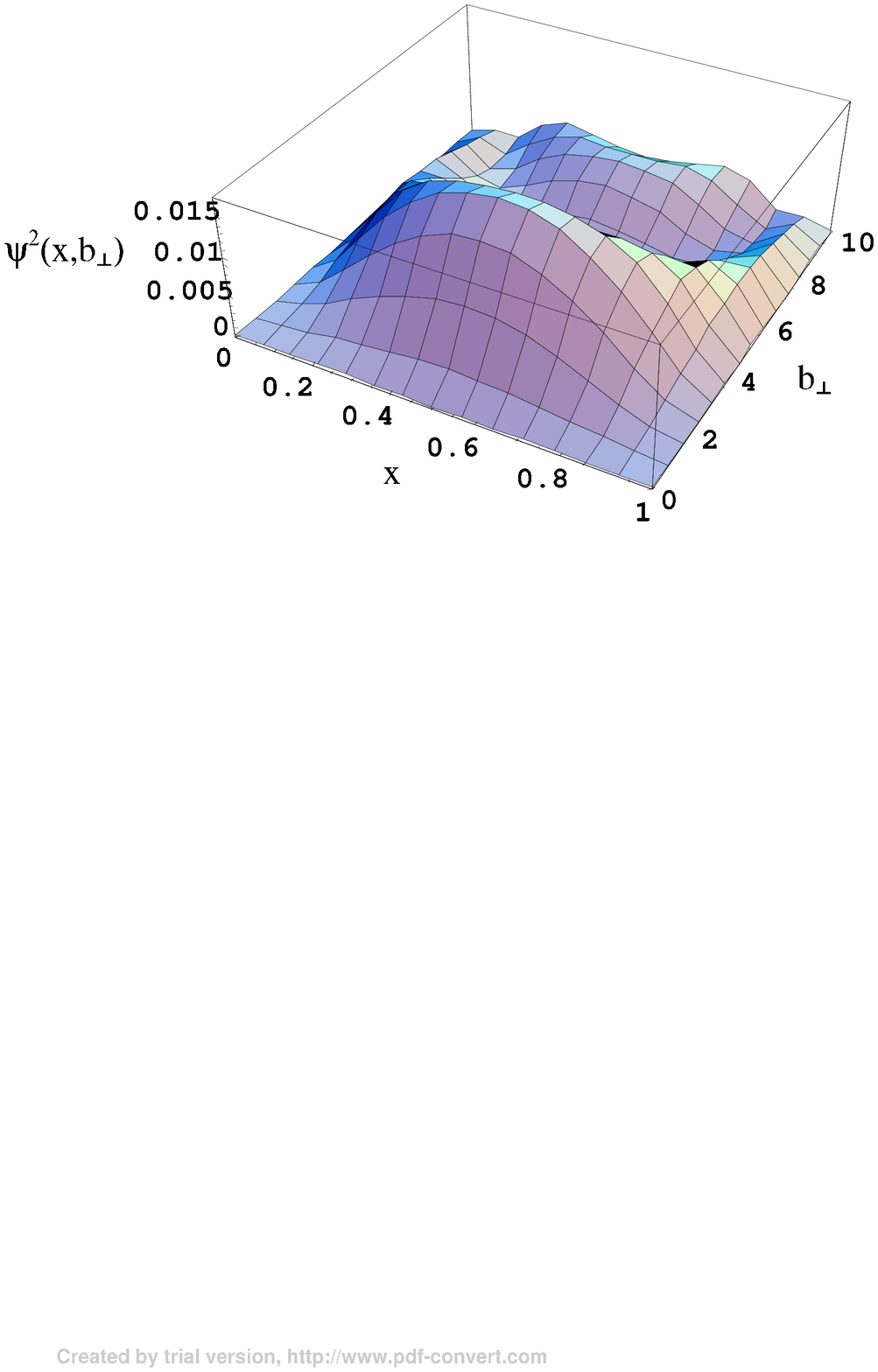}
\caption{\label{fig6} Impact parameter dependent parton distribution in the
holographic model. ~{\it LHS} is for the ground state $L=0, k=1$ and {\it RHS}
is for the first exited state $L=1, k=1$. $\Lambda_{QCD}=0.32$~ GeV and
$b_\perp$ is given in ${\mathrm{GeV}}^{-1}$.}
\end{figure}

Taking a Fourier transform (FT) of the GPDs with respect to the transverse
momentum transfer $\Delta_\perp^2$, one gets the ipdpdfs. In figs 3 and 4, we
have plotted the ipdpdfs $q(x,b_\perp)$ and $e(x,b_\perp)$ defined as in eq.
(\ref{ipd}) for models 1 and 2 respectively. The ipdpdfs show the same
qualitative behaviour in both models, which is expected as the GPDs
themselves behave in a similar manner. In other words, qualitatively, 
parton distributions
in the impact parameter space are not that sensitive to the large $k_\perp$ 
behaviour of the LFWFs. The peak is shifted to larger $x$ in model 1. 
$q(x,b_\perp)$ goes to zero at $x=0$ and $x=1$. In fact, in
the limit $x \to 1 $ the active quark with momentum fraction $x$ 
is very close to the transverse center of momentum, and the transverse width
of the ipdpdf vanishes \cite{bur}. 
The peak in $x$ decreases as $b_\perp$ increases, thus  
the transverse profile in $b_\perp$ is peaked at $\mid b_\perp \mid =0$ and
falls away further from it. The  ipdpdf with the nucleon helicity flip,
$e(x,b_\perp)$ is especially interesting, as for a transversely polarized
target, it measures the distortion of the parton distribution in the
transverse plane. In addition, the second moment of $e(x, b_\perp)$ gives
how the distribution of momentum of quarks of a particular flavour in the
transverse plane changes when the nucleon is not polarized longitudinally. 
This receives contribution from quarks with nonzero orbital angular momentum
and is shown to be connected with  Sivers effect. $e(x,b_\perp)$ is peaked
at $x=0.5$ in both models and the height of the peak decreases as $\mid
b_\perp \mid $ increases. In Fig. 5, we have shown in ipdpdfs $q(x,b_\perp)$ in
both models as a function of $\mid b_\perp \mid $ for fixed values of $x$.
It is interesting to see that in both models, there is a primary peak at
$\mid b_\perp \mid =0$ and then there are several secondary peaks. Positions
of these secondary maxima are independent of $x$. It is to be noted that in
model 2, the curve for $x=0.5$ lies above the one  for $x=0.8$, 
whereas in model 1, the situation is opposite. This is because in model 1,
the peak is shifted towards higher values of x than in model 2 (see figs 3
and 4).

In \cite{tera}, it has been shown that the string amplitude $\Phi(z)$ 
defined on the fifth dimension in $ADS_5$ space can be mapped to the
light-front wave functions of hadrons in physical spacetime. The holographic
variable $z$ corresponds to the impact variable $\xi$ where $\xi^2 = 
x (1-x) b_\perp^2$, $b_\perp$ being the impact parameter. The effective
$4$-dimensional Schroedinger equation for the bound states of massless
quarks and gluons exactly reproduce the ADS/CFT results and gives a realistic
description of the light quark meson and baryon spectrum. The normalized two
particle LFWF for the bound state is given by,

\be
\Psi_{L,k}(x,b_\perp)=B_{L,k}\sqrt{x(1-x)}J_L(\xi
\beta_{L,k}\Lambda_{QCD}); 
\ee 
where
$B_{L,k}=\Lambda_{QCD}\big[(-1)^L \pi J_{1+L}
(\beta_{L,k})J_{1-L}(\beta_{L,k})\big]^{-1/2}$, 
 $\beta_{L,k}$ is the $k$-th zero of Bessel function
$J_L$ and $\Lambda_{QCD}= 0.32$~ GeV.
For ground state $L=0,k=1$ and we have 
\be 
\Psi_{0,1}(x,b_\perp)=\Lambda_{QCD}\sqrt{x(1-x)}
{J_0(\xi\beta_{0,1}\Lambda_{QCD}) \over \sqrt{\pi}
J_1(\beta_{0,1})}.\label{adswf} 
\ee 
In fig. 6 we have plotted the ipdpdf for the ground state as 
well as the first exited state as functions of $x$ and $\mid b_\perp 
\mid $. One has to note that in this model $b_\perp$ cannot be very large as
 $z \le {1\over \Lambda_{QCD}}$. The primary peak of the ipdpdfs occur  
for the ground state ($L=0, k=1$) at $\mid b_\perp \mid =0$. 
In the first exited state ($L=1, k=1$) it is shifted away from 
$\mid b_\perp \mid =0$. In both cases, there is a secondary peak at a higher
value of $\mid b_\perp \mid $ for certain $x$ values. Similar secondary
peaks have been observed for the other two models considered, as we already
noted.    

%%%%%%%%%%%%%%%%%%%%%%%%%%%%%%%%%%%%%%%%%%%%%%%%%%%%%%%%%%%%%%%%%%%%%
\section{Summary and Conclusion}
%%%%%%%%%%%%%%%%%%%%%%%%%%%%%%%%%%%%%%%%%%%%%%%%%%%%%%%%%%%%%%%%%%%%%%%%

We have investigated GPDs as well as  impact parameter dependent parton
distributions at $\zeta=0$ in three different phenomenological models of 
hadron LFWF. 
We simulated models for meson-like and proton-like hadrons respectively,
starting from the two-particle LFWF of a dressed electron in a generalized
form of QED, and differentiating the denominator {\it wrt } the squared
masses. This improves the behaviour at the end points $x=0,1$ as well as
changes the $k_\perp$ behaviour. We found that the  GPDs and the ipdpdfs are
not that sensitive to the $k_\perp$ behaviour of the LFWFs. The peak of the
ipdpdfs in $x$ for fixed $b_\perp$ is shifted towards higher $x$ in the
simulated model for a meson-like hadron. We have also
investigated the ipdpdfs in the recently proposed QCD holographic model, and
shown the behaviour over a larger $b_\perp$ range. In all three models,
ipdpdfs show several secondary peaks in $b_\perp$ besides the primary one.  
%%%%%%%%%%%%%%%%%%%%%%%%%%%%%%%%%%%%%%%%%%%%%%%%%%%%%%%%%%%%%%%%%%%%%%%%%%%
\section{acknowledgements}
%%%%%%%%%%%%%%%%%%%%%%%%%%%%%%%%%%%%%%%%%%%%%%%%%%%%%%%%%%%%%%%%%%%%%%%
AM would like to thank DST (Fasttrack Scheme), Government of India 
 for financial support. 

%%%%%%%%%%%%%%%%%%%%%%%%%%%%%%%%%%%%%%%%%%%%%%%%%%%%%%%%%%%%%%%%%%%%%%%%

\end{document}